\documentclass[journal, letterletter, final, 10pt]{IEEEtran}
\usepackage[dvips]{graphics}
\usepackage[dvips]{changebar}
\usepackage{amsmath,bm}
\usepackage{rotating}
\usepackage[strings]{underscore}
\usepackage{setspace}
\usepackage{times}
\usepackage{color}
\usepackage{soul}
\usepackage{amsmath} 
\usepackage{amssymb}
\usepackage{mathrsfs}
\usepackage[strings]{underscore}
\usepackage{graphics,color}
 \usepackage[british,UKenglish,USenglish,english,american]{babel}
\usepackage{epstopdf}
\usepackage{graphicx}
\usepackage{multirow}
\usepackage{ifpdf}
\usepackage{tablefootnote}
\usepackage{multirow}
\usepackage{algorithmicx}
\usepackage{algpseudocode}
\usepackage{footnote}
\usepackage{gensymb}

\makesavenoteenv{tabular}
 \makesavenoteenv{table}

 \newcommand{\rev}{\textcolor[rgb]{0.0000,0.000,0.0000}} 
 \newcommand{\tj}{\textcolor[rgb]{0.0000,0.000,0.0000}} 
 \newcommand\norm[1]{\left\lVert#1\right\rVert}
 \newcommand{\mb}{\mathbf }
\newcommand{\Rtr}{\mathrm{tr} }
\newcommand{\tr}{\mathrm{tr} }
\newcommand{\R}{\mathrm{R} }

\newcommand{\ME}{\mathrm{E}}

\ifCLASSOPTIONcompsoc
\usepackage[caption=false,font=normalsize,labelfont=sf,textfont=sf]{subfig}
\else
\usepackage[caption=false,font=footnotesize]{subfig}
\fi

\begin{document}

\setcounter{page}{1}
\newcounter{mytempeqncnt}

\title{Linear Shrinkage Estimation of Covariance Matrices Using Low-Complexity Cross-Validation}
\author{Jun Tong, Rui Hu, Jiangtao Xi, Zhitao Xiao, Qinghua Guo and Yanguang Yu
 \thanks{J. Tong, R. Hu, J. Xi, Q. Guo and Y. Yu are with the School of Electrical, Computer and Telecommunications Engineering, University of Wollongong, Wollongong, NSW 2522, Australia. Email:  rh546@uowmail.edu.au, \{jtong, jiangtao, qguo, yanguang\}@uow.edu.au.} 
\thanks{Z. Xiao is with School of Electronic and Information Engineering, Tianjin Polytechnic University, China.}
 }

	\maketitle
	%\thispagestyle{empty}
	%\pagestyle{plain}	
%------------------------------------------------------------------------------------------------------------------------------------------------

\begin{abstract} 
Shrinkage can effectively improve the condition number and accuracy of covariance matrix estimation, especially for low-sample-support applications with the number of training samples  smaller than the dimensionality. This paper investigates parameter choice for linear shrinkage estimators. We propose data-driven, leave-one-out cross-validation (LOOCV) methods for automatically choosing the shrinkage coefficients, aiming to minimize the Frobenius norm of the estimation error. A quadratic loss is used as the prediction error for LOOCV. The resulting solutions can be found analytically or by solving optimization problems of small sizes and thus have low complexities. Our proposed methods are compared with various existing techniques. We show that the LOOCV method achieves near-oracle performance for shrinkage designs using sample covariance matrix (SCM) and several typical shrinkage targets. Furthermore, the LOOCV method provides low-complexity solutions for estimators that use general shrinkage targets, multiple targets, and/or ordinary least squares (OLS)-based covariance matrix estimation. We also show applications of our proposed techniques to several different problems in array signal processing.  
\end{abstract}
 

\section{Introduction} 

In statistical signal processing, one critical problem is to estimate the covariance matrix, which has extensive applications in correlation analysis, portfolio optimization, and various signal processing tasks in radar and communication systems \cite{Scharf91}-\cite{Du2010}. One key challenge is that when the dimensionality is large but the sample support is relatively low, the estimated covariance matrix $\mb R$, which may be obtained using a general method such as sample covariance matrix (SCM) or ordinary least squares (OLS), becomes ill-conditioned or even singular, and suffers from significant errors relative to the true covariance matrix $\bm \Sigma$. Consequently, signal processing tasks that rely on covariance matrix estimation may perform poorly or fail to apply. Regularization techniques have attracted tremendous attention recently for covariance matrix estimation. By imposing structural assumptions of the true covariance matrix $\bm \Sigma$, techniques such as banding \cite{Bickel2008a}, thresholding  \cite{Bickel2008b}, and shrinkage  \cite{Stein}-\cite{Ikeda2016} have demonstrated great potential for improving the performance of covariance matrix estimation. See \cite{Tong2014}-\cite{Fan2016} for recent surveys. 

This paper is concerned with the linear shrinkage estimation of covariance matrices. Given an estimate $\mb R$ of the covariance matrix, a linear shrinkage estimate is constructed as      
  \begin{equation}
      \label{ShrunkCov} 
               \widehat{\bm \Sigma}_{\rho, \tau} = \rho \mb R + \tau \mb T_0, 
 \end{equation} 
where $\mb T_0$ is the shrinkage target and $\rho$ and $\tau$ are nonnegative shrinkage coefficients. In general, the shrinkage target $\mb T_0$ is better-conditioned, more parsimonious or more structured, with lower variance but higher bias compared to the original estimate $\mb R$ \cite{Daniels2001}. The coefficients $\rho$ and $\tau$ are chosen to provide a good tradeoff between bias and variance, such that an estimate outperforming both $\mb R$ and $\mb T_0$ is achieved and a better approximation to the true covariance matrix $\bm \Sigma$ can be obtained. Compared to other regularized estimators such as banding and thresholding, linear shrinkage estimators can be easily designed to guarantee positive-definiteness. Such shrinkage designs have been employed in various applications which utilize covariance matrices and have demonstrated significant performance improvements. The linear shrinkage approach has also been generalized to nonlinear shrinkage estimation of covariance matrices \cite{Bartz2016, Lam2016}, and is closely related to several unitarily invariant covariance matrix estimators that shrink the eigenvalues of the SCM, such as those imposing condition number constraints on the estimate \cite{Aubry2012, Won2013}. There are also a body of studies on shrinkage estimation of precision matrix (the inverse of covariance matrix) \cite{Kourtis2012}-\cite{Bodnar2016} and on application-oriented design of shrinkage estimators. See \cite{Mestre06}-\cite{Tong2016ICASSP} for example applications in array signal processing.  

Shrinkage has a Bayes interpretation \cite{Ledoit2004, Haff1980}. The true covariance matrix $\bm \Sigma$ can be assumed to be within the neighborhoods of the shrinkage target $\mb T_0$. There can be various different approaches for constructing $\mb R$ and $\mb T_0$. For example, when a generative model about the observation exists, one may first estimate the model parameters and then construct $\mb R$ \cite{Senneret2016}. A typical example of this is linear models seen in communication systems. Furthermore, different types of shrinkage targets, not necessarily limited to identity or diagonal targets, can be used to better utilize prior knowledge. For example,  knowledge-aided space-time signal processing (KA-STAP) may set $\mb T_0$ using knowledge about the environment \cite{Stoica2008} or past covariance matrix estimates \cite{Guerci2006}. Even multiple shrinkage targets can be applied when distinct guesses about the true covariance matrix are available \cite{Lancewicki2014}. 

The choice of shrinkage coefficients  significantly influences the performance of linear shrinkage estimators. Various criteria and methods have been studied. Under the mean squared error (MSE) criterion, Ledoit and Wolf (LW) \cite{Ledoit2004} derived closed-form solutions based on asymptotic estimates of the statistics needed for finding the optimal shrinkage coefficients, where $\mb R$ and $\mb T_0$  are assumed as the SCM and identity matrix, respectively. Later the LW solution was extended for more general shrinkage targets \cite{Stoica2008, Lancewicki2014}. Chen \emph{et al}  \cite{Chen2010} assumed Gaussian distribution and proposed an oracle approximating shrinkage (OAS) estimator, which achieves near-optimal parameter choice for Gaussian data even with very low sample supports. The shrinkage coefficients determination can also be cast as a model selection problem and thus generic model selection techniques such as cross-validation (CV)  \cite{Arlot10}-\cite{Nowak97} can be applied. In general, CV splits the training samples for multiple times into disjoint subsets and then fits and assesses the models under different splits based on a properly chosen prediction loss. This has been explored, e.g., in \cite{Hoffbeck1996, Warton2008}, where the Gaussian likelihood is used as the prediction loss. 

All these data-driven techniques achieve near-optimal parameter choice when the underlying assumptions hold. However, there are also limitations to their applications: almost all existing analytical solutions to shrinkage coefficients \cite{Ledoit2004}-\cite{Chen2010}, \cite{Lancewicki2014} were derived under the assumption of SCM and certain special forms of shrinkage targets. They need to be re-designed when applied to other cases, which is generally nontrivial. The asymptotic analysis-based methods \cite{Ledoit2004, Stoica2008} may not perform well when the sample support is very low. Although the existing CV approaches \cite{Hoffbeck1996, Warton2008} have broader applications, they assume Gaussian distribution and employ grid search to determine the shrinkage coefficients. The likelihood cost of \cite{Hoffbeck1996, Warton2008} must be computed for multiple data splits and multiple candidates of shrinkage coefficients, which can be time-consuming.  
 
In this paper, we further investigate data-driven techniques that automatically tune the linear shrinkage coefficients using leave-one-out cross-validation (LOOCV). We choose a simple quadratic loss as the prediction loss for LOOCV, and derive analytical and computationally efficient solutions. The solutions do not need to specify the distribution of the data. Furthermore, the LOOCV treatment is applicable to different covariance matrix estimators including the SCM- and ordinary least squares (OLS)-based schemes. It can be used together with general shrinkage targets and can also be easily extended to incorporate multiple shrinkage targets.  The numerical examples show that the proposed method can achieve oracle-approximating performance for covariance matrix estimation and can improve the performance of several array signal processing schemes.     

The remainder of the paper is organized as follows. In Section 2, we present computationally efficient LOOCV methods for choosing the linear shrinkage coefficients for both SCM- and OLS-based covariance matrix estimators and also compare the proposed LOOCV methods with several  existing methods which have attracted considerable attentions recently. In Section 3, we extend our results for multi-target shrinkage. Section 4 reports numerical examples, and finally Section 5 gives conclusions. 

\section{LOOCV Choice of Linear Shrinkage Coefficients}

This paper deals with the estimation of covariance matrices of zero-mean signals whose fourth-order moments exist. We study the LOOCV choice of the shrinkage coefficients for the linear shrinkage covariance matrix estimator (\ref{ShrunkCov}), i.e., $\widehat{\bm \Sigma}_{\rho, \tau} = \rho \mb R + \tau \mb T_0$. The following assumptions are made:   
\begin{enumerate} 
\item The true covariance matrix $\bm \Sigma$, the estimated covariance matrix $\mb R$, and the shrinkage target $\mb T_0$ are all Hermitian and positive-semidefinite (PSD).   
\item $T$ independent, identically distributed (i.i.d.) samples $\{\mb y_t\}$ of the signal are available.   
\item The shrinkage coefficients are nonnegative, i.e.,   
		\begin{equation}
	  	  \label{constraints}
	  		    \rho \ge 0, \quad \tau \ge 0.    	
		\end{equation}    
\end{enumerate}
Assumption 3 follows the treatments in \cite{Ledoit2004}-\cite{Chen2010} and is sufficient but not necessary to guarantee that the shrinkage estimate $\widehat{\bm \Sigma}_{\rho, \tau}$ is PSD when Assumption 1 holds\footnote{Imposing Assumption 3 may introduce performance loss. Alternatively, one may remove the constraint $\rho \ge 0, \tau \ge 0$ and impose a constraint that $\widehat{\bm \Sigma}_{\rho, \tau}$ is PSD, similar to a treatment in \cite{Du2010}.}.  Two classes of shrinkage targets will be considered in this paper. One is constructed independent of the training samples $\{\mb y_t\}$ for generating $\mb R$, similarly to the knowledge-aided targets as considered in \cite{Stoica2008}. The other is constructed from $\{\mb y_t\}$, but is highly structured with significantly fewer free parameters as compared to $\mb R$. Examples of the second class include those constructed using only the diagonal entries of $\mb R$ \cite{Chen2010, Senneret2016} and the Toeplitz approximations of $\mb R$ \cite{Lancewicki2014}.

\subsection{Oracle Choice }
\label{OracleST} 
Different criteria may be used for evaluating the covariance matrix estimators. In this paper, we use the squared Frobenius norm of the estimation error as the performance measure. Given $\bm \Sigma$, $\mb R$ and $\mb T_0$, the oracle shrinkage coefficients minimize    
 \begin{eqnarray}
 \label{LOracle}   
 J_{\rm O}(\rho, \tau)  = || \widehat{\bm \Sigma}_{\rho, \tau}  - \bm \Sigma||_F^2  = || \rho \mb R + \tau \mb T_0  - \bm \Sigma ||_F^2,  
 \end{eqnarray} 
where $||\cdot||_F$ denotes the Frobenius norm. The cost function in (\ref{LOracle}) can then be rewritten as a quadratic function of the shrinkage coefficients:  
	  \begin{equation} 
		\label{OracleCost}
			 J_{\rm O}(\rho, \tau) = \left[ \!\!
	   \begin{array}{c} 
	      \rho   \\  
	        \tau     
	    \end{array}  
	\!\! \right]^T  \mb A_{\mathrm{O}}  
	  \left[ \!\!
	   \begin{array}{c} 
	      \rho   \\  
	        \tau     
	    \end{array}  
	\!\! \right] 
		 - 2 \left[ \!\!
	   \begin{array}{c} 
	      \rho   \\  
	        \tau     
	    \end{array}  
	\!\! \right]^T  \mb b_{\mathrm{O}}  
	 + \tr(\bm \Sigma^2),
	 \end{equation} 
\begin{equation} \label{ASTO}
\mb A_{\mathrm{O}} = \left[ \! \! 
	   \begin{array}{cc} 
	      \tr(\mb R^2) &\!\!   \Rtr (\mb R \mb T_0)   \\ 
	      \Rtr(\mb R \mb T_0)  &\!\! \tr ( \mb T_0^2)  
	    \end{array}  
	\! \! \right],  
\end{equation} 
\begin{equation} \label{bSTO} 
\mb b_{\mathrm{O}} =  \left[ \!\!
	   \begin{array}{c} 
	        \Rtr(\mb R \bm \Sigma)   \\  
	        \Rtr( \mb T_0 \bm \Sigma)     
	    \end{array}  
	\!\! \right],  
\end{equation}
where $\tr(\cdot)$ denotes the trace of a matrix. As $\mb A_{\mathrm{O}}$ is positive-definite, we can find the minimizer of $J_{\rm O}(\rho, \tau)$ by solving the above bivariate convex optimization problem. We can also apply the Karush-Kuhn-Tucker (KKT) conditions to find the solution analytically. From (\ref{OracleCost}), letting $\frac{ J_{\rm O}(\rho, \tau)}{\partial \rho} = \frac{ J_{\rm O}(\rho, \tau)}{\partial \tau} =0$ leads to 
\begin{equation} 
		\label{OracleDir1} 
			 \frac{    \tr(\mb R^2)  } { \Rtr(\mb R \bm \Sigma) }    \rho + \frac{ \Rtr (\mb R \mb T_0) } {\Rtr(\mb R \bm \Sigma)} \tau = 1,  
\end{equation} 
\begin{equation} 
		\label{OracleDir2} 
			\frac{ \Rtr (\mb R \mb T_0)}{ \Rtr( \mb T_0 \bm \Sigma)}  \rho + \frac{\tr ( \mb T_0^2)}{ \Rtr( \mb T_0 \bm \Sigma)}  \tau =1.   
\end{equation}
The oracle shrinkage coefficients can be obtained by solving (\ref{OracleDir1}) and (\ref{OracleDir2}): 		\begin{equation} 
		    \label{OracleShrunkCov}
				\left[ \!
				   \begin{array}{c} 
				      \rho^\star_{\mathrm O} \\  
				      \tau^\star_{\mathrm O} 
				    \end{array}   
				\! \right]  
				=\mb A_{\mathrm{O}}^{-1}     
				\mb b_{\mathrm{O}}.    
		\end{equation}
Note that (\ref{OracleShrunkCov}) may produce negative shrinkage coefficients, which may not lead to a positive-definite estimate of the covariance matrix. In this case, we clip the negative coefficient to zero and then find the other coefficient using (\ref{OracleDir1}) or (\ref{OracleDir2}) to guarantee the positive definiteness, for $\tau=0$ or $\rho=0$, respectively. This treatment is similar to \cite{Ledoit2004}-\cite{Du2010} and provides a suboptimal yet simple solution. The oracle estimator requires knowledge of $\bm \Sigma$, which is unavailable in real applications, but the result serves as an upper bound of the performance given the linear shrinkage structure. 

\subsection{LOOCV Choice for General Cases} 
\label{LOOCVST}

\tj{ 
Let $\widehat{\bm \Sigma}$ denote a positive-definite, Hermitian matrix. It can be easily verified that the following cost      
\begin{equation} \label{JSSigma} 
	J_{\rm S}(\widehat{\bm \Sigma}) \triangleq \mathrm{E}[|| \widehat{\bm \Sigma} - \mb y \mb y^\dag||_F^2]  
\end{equation}   
is minimized when $\widehat{\bm \Sigma}= \bm \Sigma$, where the expectation is taken over $\mb y$. In this paper, we apply LOOCV \cite{Arlot10} to produce an estimate of $J_{\rm S}(\widehat{\bm \Sigma})$ as the proxy for measuring the accuracy of $\widehat{\bm \Sigma}$, based on which the shrinkage coefficients can be selected.} With the LOOCV method, the length-$T$ training data $\mb Y=[\mb y_1, \mb y_2, \cdots, \mb y_T]$ is repeatedly split into two sets with respect to time. For the $t$-th split, where $1\le t \le T$,  $T-1$ samples in $\mb Y_t$  (with the $t$-th column $\mb y_t$ omitted from $\mb Y$) are used for producing a covariance matrix estimate $\mb R_t$ and the remaining sample $\mb y_t$ is spared for parameter validation. In total, $T$ splits of the training data $\mb Y$ are used and all the training samples are used for validation once. Assuming shrinkage estimation with given shrinkage coefficients $(\rho, \tau)$, we construct from each $\mb Y_t$ a shrinkage covariance matrix estimator as   
	 \begin{equation}
	 	\label{ShrunkCovi} 
					\widehat{\bm \Sigma}_{t, \rho, \tau} =  \rho \mb R_t  + \tau \mb T_0. 
	 \end{equation} 
We propose to use the following LOOCV cost function   
	 \begin{align}
	 	\label{LOOCVCost}      
                J_{\rm CV} (\rho, \tau) & =\frac{1}{T} \sum_{t=1}^T 
	 	 ||
	 	 \widehat{\bm \Sigma}_{t, \rho, \tau} -  \mb y_t \mb y_t^\dag 
	 	   ||_F^2 \\
                    &= \frac{1}{T} \sum_{t=1}^T 
	 	 ||
	 	  \rho \mb R_t   + \tau \mb T_0  -  \mb y_t \mb y_t^\dag  
	 	   ||_F^2    
	 \end{align} 
to approximate the cost in (\ref{JSSigma}) when $\widehat{\bm \Sigma}$ is chosen as $\widehat{\bm \Sigma}_{t, \rho, \tau}$. For notational simplicity, define 
	 \begin{equation} 
	  	\mb S_t \triangleq  \mb y_t \mb y_t^\dag. 
	 \end{equation}
After some manipulations, the above cost function can be written similarly to (\ref{OracleCost}) as   
  \begin{equation} 
   \label{JoneCost}
	 J_{\rm CV} (\rho, \tau) = \left[ \!\!
	   \begin{array}{c} 
	      \rho   \\  
	        \tau     
	    \end{array}  
	\!\! \right]^T  \mb A_{\mathrm{CV}}  
	  \left[ \!\!
	   \begin{array}{c} 
	      \rho   \\  
	        \tau     
	    \end{array}  
	\!\! \right] 
		 - 2 \left[ \!\!
	   \begin{array}{c} 
	      \rho   \\  
	        \tau     
	    \end{array}  
	\!\! \right]^T  \mb b_{\mathrm{CV}}  
	 +\frac{1}{T} \sum_{t=1}^T \tr(\mb S_t^2), 
	 \end{equation} 
where 
  \begin{equation} \label{ACV}
	\mb A_{\mathrm{CV}} =     \left[\!\!
   \begin{array}{cc} 
 \frac{1}{T}  \sum\limits_{t=1}^T \tr(\mb R_t^2) & \!\!\!\!\!\! \frac{1}{T} \sum\limits_{t=1}^T \Rtr(  \mb R_t \mb T_0) \\ 
     \frac{1}{T} \sum\limits_{t=1}^T \Rtr(  \mb R_t \mb T_0)  &    \tr(  \mb T_0^2)   
    \end{array}   \!\! 
\right],
 \end{equation} 
  \begin{equation} \label{bCV}
\mb b_{\mathrm{CV}}  = \left[ \!\!
   \begin{array}{c} 
            \frac{1}{T} \sum\limits_{t=1}^T   \Rtr( \mb R_t \mb S_t)   \\  
            \frac{1}{T}  \sum\limits_{t=1}^T  \Rtr(  \mb T_0 \mb S_t)      
    \end{array}   \!\! 
\right].   
 \end{equation} 
The shrinkage coefficients can then be found by solving the above bivariate, constant-coefficient quadratic program. Analytical solutions can be obtained under different conditions, as shown below.

\subsubsection{Unconstrained shrinkage}

For unconstrained $(\rho, \tau)$,  
setting the partial derivatives$\frac{\partial  J_{\rm CV} (\rho, \tau)}{\partial \rho} =\frac{\partial  J_{\rm CV} (\rho, \tau)}{\partial \tau}  = 0$ yields  	
\begin{equation} 
		\label{TauRhoCV}
	 \frac{\sum\limits_{t=1}^T \tr(\mb R_t^2)} {\sum\limits_{t=1}^T   \Rtr( \mb R_t \mb S_t)}    \rho   +   
	 \frac{\sum\limits_{t=1}^T \Rtr(  \mb R_t \mb T_0)}{\sum\limits_{t=1}^T   \Rtr( \mb R_t \mb S_t)}  \tau   =  1,       
		\end{equation} 
      \begin{equation} 
      	\label{TauRhoCV2} 
      	 \frac{\sum\limits_{t=1}^T \Rtr(  \mb R_t \mb T_0)  }{ \sum \limits_{t=1}^T  \Rtr(  \mb T_0 \mb S_t)} \rho  +	\frac{T \tr(  \mb T_0^2)}{ \sum\limits_{t=1}^T  \Rtr(  \mb T_0 \mb S_t)} \tau =1.
		\end{equation}  
Solving (\ref{TauRhoCV}) and (\ref{TauRhoCV2}) produces the unconstrained solution 
 \begin{equation}  \label{UncOpt}   
\left[\!\!
   \begin{array}{c} 
      \rho^\star_{\rm CV} \\  
      \tau^\star_{\rm CV}  
    \end{array}  
\!\! \right]    
=\mb A_{\mathrm{CV}}  ^{-1} \mb b_{\mathrm{CV}}.             
\end{equation} 
We choose (\ref{UncOpt}) as the optimal shrinkage coefficients if both $\rho^\star_{\rm CV}$ and $\tau^\star_{\rm CV}$ are nonnegative. 
Otherwise, we consider the optimal choices on the boundary of $\rho \ge 0, \tau \ge 0$ specified by (\ref{TauRhoCV}) or (\ref{TauRhoCV2}) for $\tau=0$ or $\rho=0$ as      
 \begin{equation}  \label{zerotaucv}
         \rho_{\rm CV}^\star =  \frac{\sum\limits_{t=1}^T \Rtr(  \mb R_t \mb S_t)}{\sum\limits_{t=1}^T \tr(\mb R_t^2)}, \quad \tau^\star_{\rm CV}=0,  
\end{equation} 
or 
   \begin{equation}   \label{zerorhocv}
         \rho_{\rm CV}^\star =0,\quad \tau_{\rm CV}^\star = \frac{ \sum\limits_{t=1}^T  \Rtr(  \mb T_0 \mb S_t)}{ T  \mathrm{tr}(  \mb T_0^2)}.       
 \end{equation}

\subsubsection{Constrained shrinkage}

For the more parsimonious design using convex linear combination, the following constraint is imposed: 
\begin{equation} \label{Convex}
	\rho = 1- \tau. 
\end{equation} 
By plugging (\ref{Convex}) into the cost function (\ref{LOOCVCost}) and taking the minimizer, we can also easily find the optimal shrinkage coefficients using   
\begin{equation}  \label{ConOpt}
 \rho_{\rm CV}^\star = 
\frac
{
\sum\limits_{t=1}^T \left(\tr(\mb T_0^2 ) -  \Rtr (\mb R_t \mb T_0) - \Rtr (\mb T_0 \mb S_t) +  \Rtr ( \mb S_t \mb R_t) \right) 
}
{
\sum\limits_{t=1}^T \left( \tr(\mb R_t^2 ) -2  \Rtr (\mb R_t \mb T_0) + \tr( \mb T_0^2      ) \right)
}.
\end{equation} 
In case a negative shrinkage coefficient is produced, we set it to zero and let the other be one according to (\ref{Convex}). Note that although the closed-form solution involves multiple matrix operations, the quantities involved need to be computed only once. Furthermore, the computational complexity may be greatly reduced given a specific method of covariance matrix estimation. In the following two subsections, we will show the simplified solutions for SCM- and OLS-based covariance matrix estimation.

\subsection{LOOCV Choice for SCM-Based Estimation} 
      We consider in this subsection that $\mb R$ is the SCM estimate of $\bm \Sigma$. In this case,  
     \begin{eqnarray}
     		\label{SCM}     
     		 \mb R = \frac{1}{T} \sum_{t=1}^T \mb y_t \mb y_t^\dag =  \frac{1}{T} \sum_{t=1}^T \mb R_t = \frac{1}{T} \sum_{t=1}^T \mb S_t,  
      \end{eqnarray} 
which is a sufficient statistic for Gaussian-distributed data when the mean vector is the zero vector. For the $t$-th split, the SCM constructed from all the samples except the $t$-th is 
     \begin{eqnarray}   
	\mb R_t = \frac{1}{T-1} \sum_{j \ne t} \mb y_j \mb y_j^\dag = \frac{T}{T-1} \mb R - \frac{1}{T-1}  \mb S_t.     
      \end{eqnarray} 
We can then verify the following expressions for quickly computing the relevant quantities in (\ref{ACV}) and (\ref{bCV}):   
\begin{equation} \label{RtTwoAve}
  \frac{1}{T} \sum\limits_{t=1}^T \tr(\mb R_t ^2 ) = \frac{T (T-2)}{(T-1)^2}  \tr(\mb R^2) - \frac{1}{T(T-1)^2} \sum_{t=1}^T ||\mb y_t||_F^4,   
\end{equation} 
\begin{equation} 
   \frac{1}{T} \sum\limits_{t=1}^T \Rtr(\mb R_t \mb S_t)   = \frac{T}{T-1}\tr(\mb R^2) - \frac{1}{T(T-1)} \sum_{t=1}^T ||\mb y_t||_F^4,   
\end{equation} 
\begin{eqnarray}   
\frac{1}{T} \sum\limits_{t=1}^T  \Rtr( \mb R_t \mb T_0) =     \Rtr ( \mb R\mb T_0  ),   
\end{eqnarray} 
\begin{eqnarray}    \label{trStTo}
\frac{1}{T} \sum\limits_{t=1}^T  \Rtr( \mb S_t \mb T_0) =   \Rtr\left( \mb R\mb T_0  \right).   
\end{eqnarray} 
Plugging these into (\ref{ACV}) and (\ref{bCV}) and after some manipulations, we can rewrite the \rev{LOOCV} cost function (\ref{JoneCost}) as 
\begin{align} \label{Jonetwo}\nonumber 
J_{\rm CV}(\rho, \tau)&= \frac{\rho T(\rho T -2\rho - 2 T + 2) }{(T-1)^2} \tr(\mb R^2)   \\\nonumber
&+ 2 \tau(\rho-1) \Rtr(\mb R  \mb T_0)  + \tau^2 \tr( \mb T_0^2 ) \\  
&+ \frac{1}{T} \left( \frac{\rho}{T-1} +1 \right)^2 \sum_{t=1}^T \norm{\mb y_t}_F^4 .
\end{align}
The optimal shrinkage coefficients can then be obtained analytically from the SCM $\mb R$, the shrinkage target $\mb T_0$, and the training samples $\{\mb y_t\}$, as discussed below.

\subsubsection{Unconstrained shrinkage}  

It can be verified from (\ref{TauRhoCV2}) and (\ref{trStTo}) that the optimal shrinkage coefficients (ignoring the nonnegative constraint $\rho \ge 0, \tau \ge 0$) satisfy 
	\begin{equation}  \label{tauast}
			\tau   =(1-\rho )  \frac{ \Rtr(\mb R \mb T_0) }{\tr(\mb T_0^2)}. 
	\end{equation} 
The closed-form solution to $\rho$ is given by
	\begin{equation}  \label{rhoast} 
	\rho^\star_{\rm CV, SCM}  = 
	\frac{ 
	          \frac{T \tr(\mb R^2) }{T-1} 
	        - \frac{ (\Rtr(\mb R \mb T_0))^2 }{ \Rtr(\mb T_0^2)} 
	        - \frac{\sum\limits_{t=1}^T \norm{\mb y_t}_F^4}{T(T-1)}  
	       }
	       {
	        \frac{(T^2-2T) \tr(\mb R^2) }{(T-1)^2} 
	        - \frac{ (\Rtr(\mb R \mb T_0))^2 }{\tr(\mb T_0^2)} 
	        + \frac{\sum\limits_{t=1}^T \norm{\mb y_t}_F^4}{T(T-1)^2}   
	       }. 
	\end{equation}  
In case $\rho^\star_{\rm CV, SCM}  > 1 $ or $\rho^\star_{\rm CV, SCM}  < 0$, we apply (\ref{zerotaucv}) or (\ref{zerorhocv}), respectively, to determine the solution, using the  expressions in (\ref{RtTwoAve})-(\ref{trStTo}).  

Note that for the typical shrinkage target $\mb T_0 = \frac{\tr(\mb R)}{N} \mb I$, (\ref{tauast}) results in $\tau^\star_{\rm CV, SCM} =1-\rho^\star_{\rm CV, SCM} $. This provides another justification for the convex linear combination design with an identity target, which has been widely adopted in the literature, e.g., \cite{Chen2010}. This also shows that for such a special target the unconstrained solution is equivalent to the constrained solution, which does not hold for more general  shrinkage targets. 

\subsubsection{Constrained shrinkage}  
For the widely considered convex linear combination with constraint 
$
\rho + \tau = 1, 
$
the optimal $\rho$ (ignoring the nonnegative constraint) is computed as 
 \begin{equation}
\label{rhoastconvex}  {
\rho^\star_{\rm CV, SCM}  = \frac{
		\frac{T\tr (\mb R^2) }{T-1}  - 2 \Rtr(\mb R \mb T_0) + 
		\tr(\mb T_0^2) - \frac{\sum\limits_{t=1}^T \norm{\mb y_t}_F^4}{T(T-1)} 
        }
        {
        \frac{(T^2-2T) \tr(\mb R^2) }{(T-1)^2}  - 2 \tr(\mb R \mb T_0) + 
		\tr(\mb T_0^2) + \frac{\sum\limits_{t=1}^T \norm{\mb y_t}_F^4}{T(T-1)^2}}.}  
\end{equation} 
Similarly, in case a negative shrinkage coefficient is obtained, we set it to zero and let the other be one.  

The above results show that the optimal shrinkage coefficients for the covariance matrix estimate (\ref{ShrunkCov}) can be computed directly from the samples and shrinkage target, without the need of specifying any user parameters. The constrained shrinkage design may lead to certain performance loss as compared to the unconstrained one. 

\subsection{LOOCV Choice for OLS-Based Covariance Estimation} 
\label{LOOCVOLS}
One advantage of the LOOCV method is that it can be applied to different covariance matrix estimators. In this subsection, we discuss the LOOCV method for OLS-based covariance matrix estimation. Note that most existing analytical solutions for choosing the shrinkage coefficients assume SCM and specific shrinkage targets and need to be re-derived for general cases. Also, in contrast to general applications of LOOCV which require a grid search of the parameters and thus a high computational complexity, we have shown that for SCM, fast analytical solutions can be obtained for choosing the shrinkage coefficients. This will also be the case for the OLS-based covariance matrix estimation. 

Consider the case with observation $\mb y \in \mathbb{C}^{N }$ modeled as 
		       \begin{equation} 
		               \label{linearmodel} 
		          		\mb y = \mb H  \mb x + \mb z,           
		          \end{equation} 
where $\mb H  \in \mathbb{C}^{N \times M}$ is a deterministic channel matrix and $\mb z  \in \mathbb{C}^{N}$ a zero-mean, white noise with covariance matrix $\sigma^2 \mb I$, which is uncorrelated with the zero-mean input signal $\mb x  \in \mathbb{C}^{M }$ with covariance matrix $\mb I$. If both training samples of $\mb x$ and $\mb y$ are known, we may first estimate the channel matrix $\mb H$ and the covariance matrix of the noise $\mb z$ using the ordinary least squares (OLS) approach. Let the block of training data be $(\mb X, \mb Y)$, where the input signal $\mb X$ can be designed to have certain properties such as being orthogonal. The OLS estimates of the channel matrix and noise variance are then  obtained as 
 %  \begin{equation}  \label{OLSH} 
     %     	\widehat{\mb H} = \mb Y \mb X^\dag \left( \mb X \mb X^\dag \right)^{-1},   
    %\end{equation} 
    \begin{align}    \label{OLSH} 
\widehat{\mb H} &= \mb Y \mb X^\dag \left( \mb X \mb X^\dag \right)^{-1},\\ \nonumber
	       \widehat{\sigma^2}  &= \frac{1}{TN} \norm{\mb Y - \widehat{\mb H} \mb X}_F^2 \\ \nonumber 
	&= \frac{1}{TN} \tr\left( (\mb Y - \widehat{\mb H} \mb X)(\mb Y - \widehat{\mb H} \mb X)^\dag\right)  \\ 
	&= \frac{1}{TN} \tr\left( \mb Y (\mb I  - \mb X^\dag (\mb X \mb X^\dag)^{-1} \mb X)   \mb Y^\dag \right),    
\end{align}
where $\widehat{(\cdot)}$ denotes the estimate of a quantity. In this case, the covariance matrix of $\mb y$ can be estimated as  
\begin{equation} \label{ROLS}
	\mb R = \widehat{\mb H}  \widehat{\mb H}^\dag  + \widehat{\sigma^2}  \mb I.  
\end{equation}  
Such OLS-based covariance matrix estimation may be useful for designing signal estimation schemes in wireless communications. We can apply the linear shrinkage design (\ref{ShrunkCov}) to enhance its accuracy and apply the LOOCV method (\ref{LOOCVCost}) to choose the shrinkage coefficients. Note that in this case, in the $t$-th split, we generate the covariance matrix estimate $\mb R_t$ by applying the OLS estimate to the leave-one-out samples $(\mb X_t, \mb Y_t)$ which are the subset of $(\mb X, \mb Y)$ with the pair $(\mb x_t, \mb y_t)$ omitted. The LOOCV cost is the same as (\ref{JoneCost}). In this case, the leave-one-out estimate of the covariance matrix for the $t$-th data split is   
 \begin{eqnarray}   
	\mb R_t=   \widehat{\mb H}_t \widehat{\mb H}_t^\dag  + \widehat{\sigma_t^2} \mb I, 
\end{eqnarray}  
where $\widehat{\mb H}_t$ and $\widehat{\sigma_t^2}$ denote the channel matrix and noise variance estimated from $(\mb X_t, \mb Y_t)$, respectively. A direct computation of (\ref{ACV}) and (\ref{bCV}) for evaluating the LOOCV cost performs OLS estimation for $T$ times, which incurs significant complexity. The complexity can be greatly reduced by observing that the leave-one-out OLS estimate of the channel matrix is related to the OLS channel matrix estimate $\widehat{\mb H}$ in (\ref{OLSH}) by a rank-one update:  
	\begin{eqnarray} 
			\label{HiHOriginal}   
      			\widehat{\mb H}_t=\mb Y_t \mb X_t^\dag \left( \mb X_t \mb X_t^\dag \right)^{-1} =  \widehat{\mb H} -\mb e_t \mb f_t^\dag,     
		\end{eqnarray}
 where 
       \begin{equation}
        	\mb e_t \triangleq  \mb y_t - \widehat{\mb H} \mb x_t,  
        \end{equation}
	\begin{equation} 
        	\mb f_t \triangleq  \frac{ 1}{1-  \Phi_{t}}  (\mb X \mb X^\dag)^{-1} \mb x_t. 
        \end{equation} 
In the above,  
	\begin{equation}
		\Phi_{t} \triangleq \mb x_t^\dag (\mb X \mb X^\dag)^{-1} \mb x_t 
	\end{equation} 
is the $t$-th diagonal entry of 
	\begin{equation}
		\bm \Phi = \mb X^\dag (\mb X \mb X^\dag)^{-1} \mb X.   
	\end{equation} 
Similarly, the leave-one-out estimate of the noise variance can be updated as
\begin{align} \nonumber    
	\widehat{\sigma_t^2} &= \frac{1}{N(T-1)} \tr\left( (\mb Y_t - \widehat{\mb H}_t \mb X_t)(\mb Y_t - \widehat{\mb H}_t \mb X_t)^\dag\right)  \\
&= \widehat{\sigma^2} -  \delta_t,  
\end{align} 
where 
\begin{equation} 
 \delta_t = \frac{  \norm{\mb e_t}_F^2}{N(T-1) (1-\Phi_{t})} - \frac{ \widehat{\sigma^2}}{T-1}.   
\end{equation} 
Note that both updates can be achieved with low complexity when a few matrices are computed in advance and reused. In this way, the covariance matrix estimate can be computed as 
\begin{equation}  \label{RtOLS} 
 \mb R_t = \mb R - \delta_t \mb I - \mb e_t \bm \phi_t^\dag - \bm \psi_t \mb e_t^\dag,  
\end{equation}
where $\bm \phi_t$ and $\bm \psi_t$ are defined as 
\begin{equation}
\bm \phi_t = \widehat{\mb H} \mb f_t,  
\end{equation} 
\begin{equation}
 \bm \psi_t = \bm \phi_t - ||\mb f_t||_F^2 \mb e_t.  
\end{equation} 
This shows that the leave-one-out OLS covariance matrix estimate can be obtained from $\mb R$ by corrections involving a scaled identity matrix and two rank-one updates. 
Eqn. (\ref{RtOLS}) can be exploited to compute the closed-form LOOCV solution quickly. 
From (\ref{RtOLS}), the most involved computation for finding the solution of the optimization problem (\ref{JoneCost}) can be implemented as    
\begin{align}
\label{traceRt2OLS}   \nonumber 
\frac{1}{T} \sum_{t=1}^T \tr(\mb R_t^2) &= \tr(\mb R^2) +   \frac{N  \sum_{i=1}^T  \delta_t^2}{T}  -  \frac{2 \sum_{i=1}^T  \delta_t}{T} \tr(\mb R)   \\ \nonumber 
  & +  \frac{1}{T} \sum_{i=1}^T  ||\mb e_t||_F^2 (||\bm \phi_t||_F^2 + ||\bm \psi_t||_F^2) \\ 
&   -   \frac{2}{T} \sum_{i=1}^T \R(\mb e_t^\dag( \mb R-\delta_t \mb I) (\bm \phi_t + \bm \psi_t) - \mb e_t^\dag \bm \psi_t \mb e_t^\dag \bm \phi_t ),    
\end{align} 
where $\R(\cdot)$ denotes the real part of a scalar. When $\mb R$ is already computed, the right-hand side of (\ref{traceRt2OLS}) can be evaluated using inner products and matrix-vector products. The terms $\tr(\mb T_0^2)$ and $\Rtr(\mb T_0 \mb S_t)$ are the same as those for SCM. For the other two terms, we have 
\begin{align}
\label{traceRtStOLS}  \nonumber     
	  \frac{1}{T}  \sum_{t=1}^T \Rtr(\mb R_t \mb S_t) &=\frac{1}{T}  \Rtr \left(  \mb R \mb Y \mb Y^\dag \right)   \\
&-  \frac{1}{T}  \sum_{t=1}^T \R( \delta_t \norm{\mb y_t}_F^2  + \mb y_t^\dag \mb e_t \bm \phi_t^\dag \mb y_t + \mb y_t^\dag \bm \psi_t \mb e_t ^\dag \mb y_t),     
\end{align} 
\begin{align}
\label{traceRtTzOLS}    \nonumber   
	\frac{1}{T}  \sum_{t=1}^T \Rtr(\mb R_t \mb T_0) &=\Rtr \left( \mb R   \mb T_0 \right) -  \frac{\sum_{t=1}^T \delta_t}{T}   \tr(\mb T_0)  \\ 
 &- \frac{1}{T}  \sum_{t=1}^T \R(  \bm \phi_t^\dag \mb T_0 \mb e_t +  \mb e_t ^\dag \mb T_0 \bm \psi_t).     
\end{align} 
Note that the computational complexities of (\ref{traceRt2OLS})-(\ref{traceRtTzOLS}) are low because the major operations are matrix-vector products and inner products.

\subsection{Comparisons with Alternative Choices of Linear Shrinkage Coefficients} 
\label{Alternative} 
In the above, we have introduced LOOCV methods with analytical solutions for choosing the coefficients for linear shrinkage covariance matrix estimators.  We now discuss several alternative techniques which have received considerable attentions recently and compare them with the LOOCV methods proposed in this paper. 

In 2004, Ledoit and Wolf (LW) \cite{Ledoit2004} studied estimators that shrink SCM toward an identity target, i.e., $\mb T_0 =  \mb I$.  Such estimators do not alter the eigenvectors but shrink eigenvalues of the SCM, which is well supported by the fact that sample eigenvalues tend to be more spread than population eigenvalues. The optimal shrinkage coefficients under the MMSE criterion (\ref{LOracle}) can be written as 
\begin{equation}\label{LWEst}
\rho^\star =  \frac{ \alpha^2 }{\delta^2 }, \quad \tau^\star = \frac{  \beta^2 }{\delta^2 }  \mu,     
\end{equation} 
where the parameters  
$\mu \triangleq  \frac{ \tr(\bm \Sigma)}{N}$, $\delta^2 \triangleq \ME [  \norm{ \mb R  -\mu \mb I  }^2_F ]$, $\beta^2 = \ME[ \norm{ \bm \Sigma - \mb R  }^2_F ]$, and $\alpha^2 =  \norm{ \bm \Sigma  -\mu \mb I  }^2_F$  
depend on the true covariance matrix $\bm \Sigma$ and other unknown statistics.  
Ref. \cite{Ledoit2004} shows that $\delta^2 = \alpha^2 + \beta^2$ and proposes to approximate these quantities by their asymptotic estimates under $T\rightarrow \infty, N \rightarrow \infty$, $N/T  \rightarrow c < \infty$, as     
\begin{equation}
	\widehat{\mu} = \frac{ \tr(\mb R)}{N},  \quad  
 \widehat{\delta^2} =  \norm{ \mb R  -\widehat{\mu} \mb I  }^2_F,   
\end{equation} 
\begin{equation} \label{beta}
\widehat{\beta^2} = \min\left(\widehat{\delta^2}, \frac{1}{T^2} \sum_{t=1}^T  \norm{ \mb y_t \mb y_t^\dag - \mb R }^2_F \right),   \quad 
\widehat{\alpha^2} =   \widehat{\delta^2}  - \widehat{\beta^2},    
\end{equation} 
which can all be computed from the training samples. By substituting these into (\ref{LWEst}), estimators that significantly outperform SCM are obtained, which also approach the oracle estimators when the training length is large enough.    

The above LW estimator is extended by Stoica \emph{et al} in 2008 \cite{Stoica2008} for complex-valued signals with general shrinkage targets $\mb T_0$, with applications to knowledge-aided space-time adaptive processing (KA-STAP) in radar applications. Several estimators with similar performance are derived there. For the general linear combination (GLC) design of \cite{Stoica2008}, it is shown that the oracle shrinkage coefficients for (\ref{ShrunkCov}) satisfy 
\begin{equation}\label{rhoStoica}
   \rho^\star =1- \frac{\tau^\star}{\nu},  
\end{equation}
where 
\begin{equation} \label{StoicaSol}
\nu= \frac{\tr(\mb T_0  \mb \Sigma)}{\norm{\mb T_0}_F^2} ,   \quad 
 \tau^\star =\nu \frac{\beta^2}{\ME[\norm{\mb R -\nu \mb T_0}_F^2] }. 
\end{equation}
The quantity $\beta^2$ is estimated in the same way as (\ref{beta}), and a computationally efficient expression for $\widehat{\beta^2}$ is given by    
\begin{equation}
\widehat{\beta^2} =   \frac{1}{T^2} \sum_{t=1}^T  \norm{ \mb y_t}^4_F - \frac{1}{T} \norm{\mb R}^2_F.  
\end{equation} 
Furthermore,  $\nu$ and $\ME[||\mb R- \nu \mb T_0||_F^2]$ are estimated as 
$\widehat{\nu} =  \frac{\tr(\mb T_0  \mb R)}{||\mb T_0||_F^2}$ and $ ||\mb R- \widehat{\nu} \mb T_0||_F^2$, respectively. This leads to the result given by Eqns. (34) and (35) of \cite{Stoica2008}, which can recover the LW estimator \cite{Ledoit2004} when the identity shrinkage target $\mb T_0 = \mb I$ is assumed.

More recently, Chen \emph{et al} \cite{Chen2010} derived the oracle approximating shrinkage (OAS) estimator, which assumes SCM, real-valued Gaussian samples, and scaled identity target with 
$
	 \mb T_0 = \frac{\tr(\mb R) }{N} \mb I 
$ and 
$\rho = 1-\tau$. They first derive the oracle shrinkage coefficients for SCM obtained from i.i.d. Gaussian samples, which is determined by $N, T, \tr(\bm \Sigma)$ and $\tr(\bm \Sigma^2)$. Then, they propose an iterative procedure to approach the oracle estimator. In the iterations, $\tr(\bm \Sigma^2)$ and $\tr(\bm \Sigma)$ are estimated by $\tr(\widehat{\bm \Sigma}_j \mb R)$ and $\tr(\widehat{\bm \Sigma}_j )$, respectively, where $\widehat{\bm \Sigma}_j$ is the covariance matrix estimate at the $j$-th iteration. It is further proved that $\widehat{\bm \Sigma}_j$ converges to the OAS estimator with the following analytical expression for $\tau$:  
              \begin{equation}
             	\label{Chen}
   		        \tau_{\mathrm {OAS}}^\star  = \min\left(1, \frac{
			\left(1-\frac{2}{N} \right) \tr(\mb R^2) + (\tr(\mb R))^2
									}
			{\left(T+1-\frac{2}{N} \right) [\tr(\mb R^2) - \frac{ (\tr(\mb R))^2}{N} ]
									}
  \right).    
\end{equation}  
This approach achieves superior performance for (scaled) identity target and Gaussian data and dominates the LW estimator \cite{Ledoit2004} when $T$ is small. It was later generalized by Senneret \emph{et al} \cite{Senneret2016} to a shrinkage target chosen as the diagonal entries of the SCM. Other related techniques include \cite{Fisher2011}, which also assumes SCM, Gaussian data, and identity/diagonal shrinkage targets.

All the above techniques provide analytical solutions and achieve near-oracle performance when the underlying assumptions (e.g., large dimensionality, large size of training data, identity/diagonal shrinkage targets) hold. However, they also have limitations. A common restriction is that all these analytical solutions assume SCM and are not optimized for other types of covariance matrix estimators such as model-based estimators. In particular, the LW and GLC methods \cite{Ledoit2004, Stoica2008}, which employ asymptotic approximations, may exhibit a noticeable gap to the oracle choice when the sample support is low, which may be relevant in some applications. The OAS method \cite{Chen2010} assumed identity target, but its extensions to more general cases, e.g., with multiple/general shrinkage targets, are not trivial. By contrast, the LOOCV method proposed in this paper allows different designs and achieves near-oracle performance in general. 

Cross-validation has also been applied previously for choosing shrinkage coefficients for covariance matrix estimation. The key issues for applying this generic tool include finding appropriate predictive metrics for scoring the different estimators and fast computation schemes. 
In \cite{Hoffbeck1996, Warton2008}, the Gaussian likelihood was chosen as such a proxy. The computations with likelihood are generally involved as multiple matrix inverses/determinants are required, and a grid search is required for finding the optimal parameters. In this paper, we use the distribution-free, Frobenius norm loss in (\ref{LOOCVCost}) as the metric, which leads to analytical solutions and is computationally more tractable.

\section{Multi-Target Shrinkage}  
In Section 2, we have considered linear shrinkage designs with a single target. Multiple shrinkage targets may be used to further enhance performance, which may be obtained from a priori knowledge, e.g., a past covariance matrix estimate from older training samples or from neighboring frequencies. We can easily extend our proposed LOOCV method to multiple targets. 

\subsection{Oracle choice of shrinkage coefficients}

Consider the multi-target shrinkage design 
 \begin{equation}
	 	\label{MT} 
					\widehat{\bm \Sigma}_{ \rho, \bm \tau} =  \rho  \mb R + \sum_{k=1}^K \tau_k \mb T_k,   
	 \end{equation}  
where all the shrinkage coefficients are nonnegative to guarantee PSD covariance matrix estimates, i.e.,  
\begin{equation}
	\rho \ge 0; \qquad \tau_k \ge 0, \forall k. 
\end{equation} 
The oracle multi-target shrinkage minimizes the squared Frobenius norm of the estimation error   
	 \begin{equation}
	 	\label{OracleMT}  
	  J_{\rm O, MT} (\rho, \bm \tau)  =   
	 	   \norm{
	 	   \rho \mb R  + \sum_{k=1}^K \tau_k \mb T_k   - \bm \Sigma 
	 	   }_F^2,   
	 \end{equation}  
 which can be rewritten as   
	  \begin{equation} 
        \label{OracleCostMT}
	     J_{\rm O, MT} (\rho, \bm \tau)  = \left[ \!\!
	   \begin{array}{c} 
	      \rho   \\  
	         \bm \tau     
	    \end{array}  
	\!\! \right]^T  \mb A_{\mathrm{O, MT}}  
	  \left[ \!\!
	   \begin{array}{c} 
	      \rho   \\  
	   \bm     \tau     
	    \end{array}  
	\!\! \right] 
		 - 2 \left[ \!\!
	   \begin{array}{c} 
	      \rho   \\  
	         \bm \tau     
	    \end{array}  
	\!\! \right]^T  \mb b_{\mathrm{O, MT}}  
	 + \tr(\bm \Sigma^2), 
	 \end{equation} 
where $\bm \tau = [\tau_1, \tau_2, \cdots, \tau_K]^T$, 
\begin{equation}   \mb A_{\mathrm{O, MT}}    = \left[\!\!
   \begin{array}{cccc} 
    \tr(\mb R^2) \! \!\!&\!\!  \Rtr(\mb R \mb T_1) \!\!&\!\!  \cdots \!\!&\!\!  \Rtr(\mb R \mb T_K)    \\ 
     \Rtr(  \mb T_1\mb R ) \! \!\!&\!\!     \tr( \mb T_1^2) \!\!&\!\!  \cdots \!\!&\!\!        \Rtr(  \mb T_1 \mb T_K)    \\ 
	\vdots & \vdots & \ddots  & \vdots \\ 
   \Rtr(  \mb T_K \mb R)  \!\!&\!\!      \Rtr(  \mb T_K \mb T_1) \!\!&\!\!  \cdots \!\!&\!\!     \tr( \mb T_K^2)     
    \end{array}   
\right],   	 \end{equation}  
\begin{equation} 
 \label{MTSolt}  
 \mb b_{\mathrm{O, MT}} =     \left[\!\!
   \begin{array}{c} 
      \Rtr(  \mb R  \bm \Sigma)    \\ 
      \Rtr( \mb T_1 \bm \Sigma)   \\ 
	\vdots  \\ 
  \Rtr( \mb T_K  \bm \Sigma) 
    \end{array}   
\right].    
 \end{equation}  
The oracle shrinkage coefficients can then be obtained by solving the problem of minimizing the cost function $J_{\rm O, MT} (\rho, \bm \tau)$ of (\ref{OracleCostMT}), which is a strictly convex quadratic program (SCQP) with $K+1$ variables.

\subsection{LOOCV choice of shrinkage coefficients}

We now extend the LOOCV method in Section 2 to the multi-target shrinkage here. Following the same treatment as in Section \ref{LOOCVST}, in each split of the training data, $\mb R_t$ and $\mb S_t$ are constructed to generate and validate the covariance matrix estimate, respectively. The multiple shrinkage coefficients are chosen to minimize the LOOCV cost 
	 \begin{equation}
	 	\label{LOOCVCostMT2}  
	 	   J_{\rm CV, MT} (\rho, \bm \tau) =  \frac{1}{T} \sum_{t=1}^T 
	 	 \norm{
	 	   \rho \mb R_t  + \sum_{k=1}^K \tau_k \mb T_k   - \mb S_t
	 	   }_F^2.   
	 \end{equation}  
The above cost function can be rewritten  in a form similar to (\ref{JoneCost}) as 
\begin{align}  \nonumber 
 J_{\rm CV, MT}(\rho, \bm \tau)  &=
\left[ \!\!
	   \begin{array}{c} 
	      \rho   \\  
	         \bm \tau     
	    \end{array}  
	\!\! \right]^T  \mb A_{\mathrm{CV, MT}}  
	  \left[ \!\!
	   \begin{array}{c} 
	      \rho   \\  
	       \bm \tau     
	    \end{array}  
	\!\! \right]  - 2 \left[ \!\!
	   \begin{array}{c} 
	      \rho   \\  
	         \bm \tau     
	    \end{array}  
	\!\! \right]^T  \mb b_{\mathrm{CV, MT}}  \\
	& +\frac{1}{T}\sum_{t=1}^T \tr(\mb S_t^2)
	 \end{align}  
with 	 
\begin{equation}  \label{MTSolt}
 \mb A_{\mathrm{CV, MT}}   = \left[\!\!
   \begin{array}{cccc} 
    \frac{\sum\limits_{t=1}^T  \tr(\mb R_t^2)}{T}  \!\!\!&\!\!\!  \frac{ \sum\limits_{t=1}^T \Rtr(\mb R_t \mb T_1)}{T} \!\!\!&\!\!\!  \cdots \!\!\!&\!\!\!    \frac{\sum\limits_{t=1}^T \Rtr (\mb R_t \mb T_K)}{T}    \\ 
       \frac{\sum\limits_{t=1}^T  \Rtr( \mb T_1\mb R_t)}{T}  \!\!&\!\!     \tr( \mb T_1^2) \!\!&\!\!  \cdots \!\!&\!\!   \Rtr( \mb T_1 \mb T_K)    \\ 
	\vdots & \vdots & \ddots  & \vdots \\ 
  \frac{\sum\limits_{t=1}^T \Rtr( \mb T_K \mb R_t)}{T}  \!\!&\!\!    \Rtr(\mb T_K \mb T_1) \!\!&\!\!  \cdots \!\!&\!\!     \tr( \mb T_K^2)     
    \end{array}   
\right], 
 \end{equation}  
\begin{equation}  \label{MTSolt}
\mb b_{\mathrm{CV, MT}}=\left[\!\!
   \begin{array}{c} 
    \frac{1}{T}\sum\limits_{t=1}^T  \Rtr(\mb R_t \mb S_t)     \\ 
    \frac{1}{T}   \sum\limits_{t=1}^T \Rtr( \mb T_1 \mb S_t)   \\ 
	\vdots  \\ 
  \frac{1}{T}\sum\limits_{t=1}^T \Rtr( \mb T_K  \mb S_t) 
    \end{array}   
\right].
 \end{equation}  
The constant entries of $\mb A_{\mathrm{CV, MT}}$ and $\mb b_{\mathrm{CV, MT}}$ can be computed in the same way as for the single-target case. When $K$ is small, which is typically the case, the solution that minimizes the LOOCV cost can be found quickly using standard optimization tools. Alternatively, we may find first the global optimizer that ignores the nonnegative constraint by     
 \begin{equation}  \label{MTSolt}
\left[\!\!
   \begin{array}{c} 
      \rho^\star_{\mathrm{CV,MT}} \\  
      \bm \tau^\star_{\mathrm{CV,MT}}  
    \end{array}  
\!\! \right]    =\mb A_{\mathrm{CV, MT}} ^{-1} \mb b_{\mathrm{CV, MT}}, 
\end{equation}  
and check if the nonnegative condition is satisfied. 
If a negative shrinkage coefficient is produced, we then consider the boundaries of $\rho\ge 0, \tau_k \ge 0, k=1,2,\cdots, K$, which are equivalent to removing a certain number of shrinkage targets from the shrinkage design. The solution can be found in exactly the same way as (\ref{MTSolt}) but with fewer targets.

Similarly to the single-target case, we may also consider a constrained case, where the shrinkage targets $\{\mb T_k\}$ have the same trace as the estimated covariance matrix $\mb R$, and 
	 \begin{equation} \label{Constrained}
		\rho  + \sum_{k=1}^K \tau_k = 1.   
	 \end{equation}  
Then the LOOCV cost function can be rewritten as 
	 \begin{equation}
	 	\label{LOOCVCostMT}  
	 	 {  J_{\rm CV, MT}( \bm \tau)  =  \frac{1}{T} \sum_{t=1}^T 
	 	 \norm{
	 	  \sum_{k=1}^K     \tau_k \mb A_{kt}  + \mb B_t  }_F^2},   
	 \end{equation}   
	 where
    \begin{equation}
	 	\mb A_{kt} \triangleq \mb T_k - \mb R_t, 1\leq k\leq K, 1\leq t\leq T,    
 \end{equation}   
	\begin{equation}
	 	\mb B_{t}  \triangleq  \mb R_t  - \mb S_t, 1\leq t\leq T.      
 \end{equation}   
	 The optimal shrinkage coefficients can be found similarly as for the unconstrained case by minimizing   
	 \begin{equation}  J_{\rm CV, MT}(\bm \tau)  = 
	 \bm \tau^T  \mb A'_{\mathrm{CV, MT}}  
	 \bm \tau
		 - 2  
	         \bm \tau^T  \mb b'_{\mathrm{CV, MT}}  
	 +\frac{1}{T}\sum_{t=1}^T \tr(\mb B_t^2), 
	 \end{equation}  	 
where the entries of $\mb A'_{\mathrm{CV, MT}}$ and $\mb b'_{\mathrm{CV, MT}}$ are defined by  
\begin{equation}
[ \mb A'_{\mathrm{CV, MT}}  ]_{mn} \triangleq  \frac{1}{T}     \sum\limits_{t=1}^T \Rtr ( {\mb A}_{mt}   {\mb A}_{nt}  ),    1\leq m, n \leq K, 
 \end{equation}   
\begin{equation}
[ \mb b'_{\mathrm{CV, MT}} ]_k  \triangleq \frac{1}{T}  \sum\limits_{t=1}^T  \Rtr( \mb A_{kt}  \mb B_t),  1\leq k\leq K.   
 \end{equation}   
 These entries may also be evaluated quickly. For example, with SCM,    
\begin{equation}
[ \mb A'_{\mathrm{CV, MT}}  ]_{mn} =  \Rtr(\mb T_m \mb T_n) -\Rtr((\mb T_m + \mb T_n) \mb  R) + \eta,   
 \end{equation}   
\begin{equation}
[ \mb b'_{\mathrm{CV, MT}} ]_k  =   
 \frac{T}{T-1} \tr( \mb R^2) - \frac{1}{T(T-1)} \sum_{t=1}^T ||\mb y_t||_F^4 - \eta,  
 \end{equation}   
 where 
 \[
 \eta = \frac{1}{T} \sum_{t=1}^T \tr(\mb R_t^2)  
 \]
can be computed using (\ref{RtTwoAve}). The solution to $\bm \tau$ can be found as 
\begin{equation}
   \bm \tau^\star_{\mathrm{CV, MT}} =   \mb {A'}_{\mathrm{CV, MT}} ^{-1} \mb b'_{\mathrm{CV, MT}} 
 \end{equation}   
 if the nonnegative condition is satisfied. 
 Otherwise, find the solution in a similar way as for the unconstrained case. 
 
Note that for multi-target shrinkage, Lancewicki and Aladjem \cite{Lancewicki2014} recently introduced another method for finding the shrinkage coefficients. They assume SCM and shrinkage targets which belong to a set that can be characterized by Eqn. (21) of \cite{Lancewicki2014}. Then, they follow the Ledoit-Wolf (LW) framework \cite{Ledoit2004} to derive unbiased estimates of the unknown coefficients needed for minimizing the expectation of the cost in (\ref{OracleMT}), based on which $\{\rho, \bm \tau\}$ can be optimized. By contrast, our approach resorts to a LOOCV estimate of the cost in (\ref{JSSigma}), which does not rely on the aforementioned assumptions in \cite{Lancewicki2014}. As will be shown later, the LOOCV method can achieve similar performance as \cite{Lancewicki2014} for the shrinkage targets considered there. However, it can be applied to general estimators other than SCM and shrinkage targets which are not covered by Eqn. (21) of \cite{Lancewicki2014}, offering wider applicability.   
   \begin{figure}\centering
           \label{ARModel}
           \includegraphics[width=0.85\columnwidth]{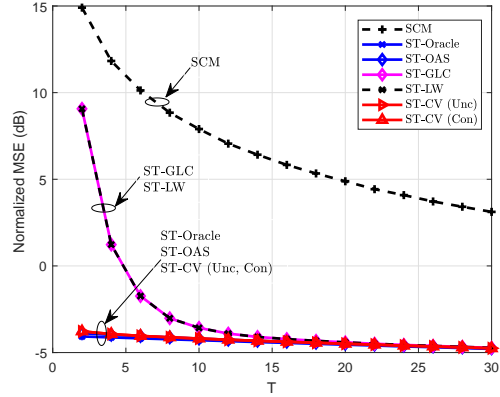}
           \caption{NMSE of single-target (ST) shrinkage estimates of an AR covariance matrix with $N=100, r=0.5$, $\mb T_0 = \frac{\tr(\mb R)} {N} \mb I$.  ``LW", ``GLC" and ``OAS" refer to the methods of \cite{Ledoit2004}, \cite{Stoica2008} and \cite{Chen2010}, respectively, which are also described in Section \ref{Alternative}; ``CV" refers to our proposed LOOCV method; ``Oracle" refers to the coefficient choice in Section \ref{OracleST}; and ``Con" and ``Unc" indicate that the constraint $\rho+\tau=1$ is imposed or not, respectively.}   
    \end{figure}     

 \section{Numerical Examples} 
In this section, we present numerical examples to demonstrate the effectiveness of the proposed shrinkage design and compare it with alternative methods. The quality of covariance matrix estimation is measured by the MSE normalized by the average of the squared Frobenius norm $||\bm \Sigma||_F^2$, i.e., 
\begin{equation} 
	\mathrm{NMSE}_{\bm \Sigma} \triangleq \frac{\mathrm{E}[ || \widehat{\bm \Sigma}_{\rho, \tau}  - \bm \Sigma||_F^2]}{\mathrm{E}[ ||  \bm \Sigma||_F^2]}.  
\end{equation}  
We show examples of covariance matrix estimation and its applications in array signal processing. We denote by $\mathcal{N}(\mu, \sigma^2)$ a real-valued Gaussian distribution with mean $\mu$ and variance $\sigma^2$.

\emph{Example 1: Shrinkage toward an identity target:}
We first consider a real-valued example with an autoregressive (AR) covariance matrix, whose $(i,j)$-th entry is given by 
 \begin{equation}
 	 \label{ARCov}  
 	 [\bm \Sigma]_{i,j} =  r^{ |i-j|},   1\leq i, j \leq N, 
 \end{equation} 
which has been widely considered for evaluating covariance matrix estimation techniques \cite{Chen2010}-\cite{Bickel2008b}. Let $ {\bm \Sigma}^{1/2}$ be the Cholesky factor of $\bm \Sigma$.  The training samples are randomly generated as $\mb y_t =  {\bm \Sigma}^{1/2} \mb n_t$, where $\mb n_t$ consists of i.i.d. entries drawn from $\mathcal{N}(0,1)$. The typical shrinkage target $\mb T_0 = \frac{\tr(\mb R)} {N} \mb I$ is considered for single-target shrinkage. Our proposed LOOCV method is compared with the widely used alternative methods \cite{Ledoit2004}-\cite{Chen2010} for choosing the shrinkage coefficients. The simulation results (averaged over $1000$ repetitions for each training length $T$) in Fig. 1 confirm that the LOOCV methods with and without the constraint $\rho+\tau=1$ produce the same results for the scaled identity target and they achieve performance almost identical to the OAS estimator \cite{Chen2010}, which was derived by assuming Gaussian data and identity target.  The LW \cite{Ledoit2004} and GLC \cite{Stoica2008} methods, which are equivalent for the scaled identity target here, do not perform well for very low sample support, but are able to approximate the oracle choice very well when more samples are available, which is consistent with the observations from \cite{Chen2010}. All of these shrinkage designs significantly outperform the SCM, confirming the effectiveness of shrinkage for covariance matrix estimation. Recall that these methods were derived using different strategies and assumptions and have different analytical solutions.

\emph{Example 2: Shrinkage toward a nondiagonal target:}
We then consider an example of the linear model given by (\ref{linearmodel}). For each training length, $1000$ random realizations of $\bm \Sigma = \mb H \mb H^\dag + \sigma^2 \mb I$ are generated and estimated through training, where $\sigma^2=0.1$. The entries of $\mb H$ are independently generated from $\mathcal{N}(0, 1)$ and then fixed for the whole training process. Given $\mb H$, $T$ training samples are generated by $\mb y = \mb H  \mb x + \mb z$, with the entries of $\mb x$ and $\mb z$ generated independently from $\mathcal{N}(0, 1)$ and $\mathcal{N}(0, \sigma^2)$, respectively. In order to demonstrate the effectiveness of the LOOCV method for general shrinkage targets, we assume a scenario where $\mb H$ is slowly time-varying and the shrinkage target $\mb T_0$ can be constructed as a well-conditioned estimate of a past covariance matrix 
\begin{equation}
 \bm \Sigma^{\rm past} =  {\mb H}^{\rm past}  {\mb H}^{\rm past \dag} + \sigma^2 \mb I,    
\end{equation}  
where  
\begin{equation}
    {\mb H}^{\rm past}  = \mb H + \bm \Delta,   
\end{equation}  
and the entries of $\bm \Delta$ are independently drawn from $\mathcal{N}(0, 0.2)$ and are fixed for each repetition. Specifically, we construct $\mb T_0$ as the shrinkage estimate of $\bm \Sigma^{\rm past}$ using SCM and the scaled identity target. This construction is similar to the knowledge-aided target considered in \cite{Stoica2008} and the resulting $\mb T_0$ is not diagonal. We assume that the numbers of samples used for estimating $\bm \Sigma$ and $\mb T_0$ are both equal to $T$. The simulation results are included in Fig. 2 for the normalized MSE. It can be seen that the LOOCV methods generally achieve near-oracle performance and outperform the GLC method. Also, the non-diagonal shrinkage target achieves better performance than the scaled identity target.  

\begin{figure}\centering
                \label{Result}
                 \includegraphics[width=0.9\columnwidth]{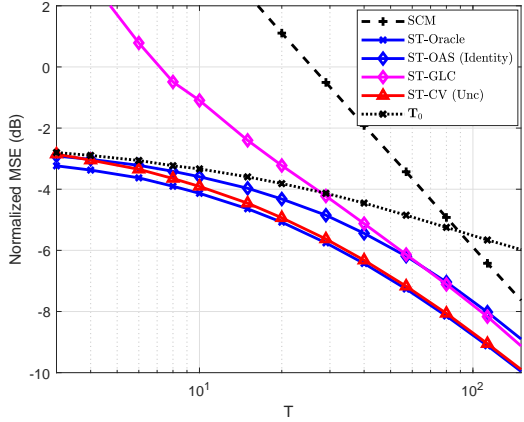}
                 \caption{NMSE of single-target (ST) shrinkage estimation of covariance matrix for the linear model (\ref{linearmodel}) with $N=50, M=50, \sigma^2=0.1$.  The non-diagonal shrinkage target is constructed from the estimate of a past covariance matrix. The result indicated by ``$\mb T_0$" corresponds to estimating $\bm \Sigma$ as $\mb T_0$. ``Identity" indicates a scaled identity shrinkage target is used instead. We can show that imposing the constraint (\ref{Convex}) leads to negligible change in performance for the proposed LOOCV approach.}    
    \end{figure}

\emph{Example 3: Shrinkage with multiple targets:}
A multi-target example is illustrated in Fig. 3. An AR covariance matrix is estimated by shrinking SCM with three targets which can be represented by Eqn. (21) of \cite{Lancewicki2014}: $\mb T_1 = \frac{\tr(\mb R)}{N}\mb I$, $\mb T_2 = \mathrm{Diag}(\mb R)$, and $\mb T_3$ is a symmetric, Toeplitz matrix which was considered in \cite{Lancewicki2014}:   
\begin{equation} \label{ModelT3}
\mb T_3 = \frac{\tr(\mb R)}{N} \mb I + \sum_{i=1}^{N-1} \frac{ \tr( \mb C_i \mb R) }{2(N-i)} \mb C_i,   
\end{equation}   
where $\mb C_i$ is a symmetric, Toeplitz matrix with unit entries on the $i$-th sub- and super-diagonals and zeros elsewhere.  It is seen that multi-target shrinkage can significantly outperform single-target shrinkage with $\mb T_0 = \frac{\tr(\mb R)}{N}\mb I$ when the number of samples is large enough. For the oracle parameter choices, the unconstrained shrinkage design, which allows a larger set of shrinkage factors to be chosen, can noticeably outperform the design constrained by (\ref{Constrained}). However, when the proposed LOOCV methods are used, the gap is significantly reduced. We can show that when the number of samples is small, using a more parsimonious design with constrained shrinkage coefficients or fewer shrinkage targets may achieve better performance.  It is seen that the multi-target shrinkage method of \cite{Lancewicki2014} (indicated by ``MT-LA" in Fig. 3) performs similarly to the LOOCV method for this example. Note that the method of \cite{Lancewicki2014} assumes SCM and shrinkage targets satisfying certain structures and does not apply directly to model-based covariance matrix estimation or more general shrinkage targets.   

  	\begin{figure}\centering
                \label{Result}
                 \includegraphics[width=0.85\columnwidth]{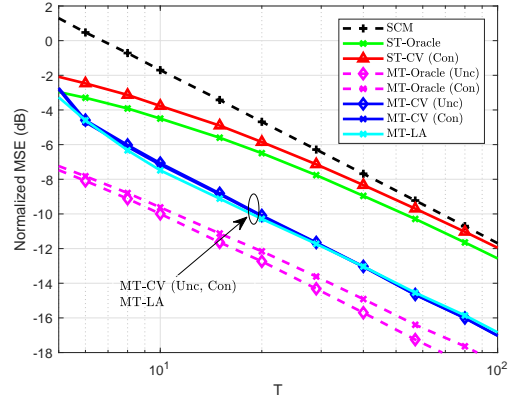}
                 \caption{MSE of covariance matrix estimation with multi-target (MT) shrinkage and LOOCV parameter choices.  AR covariance matrix with $N=50, r=0.9$ is assumed. ``LA" refers to the method proposed by Lancewicki and Aladjem \cite{Lancewicki2014}. Note that the LOOCV methods and the LA method achieve similar performance for this example. 
                 }  
    \end{figure}      

\emph{Example 4: Application to MMSE estimation of MIMO channels.} 
A potential application of the proposed technique is the design of MMSE estimator of MIMO channels. Consider a point-to-point MIMO system with $N_t$ transmitting antennas and $N_r$ receiving antennas. Let $B$ be the length of the pilot sequence. The received signal matrix during the training stage is modelled as 
\begin{equation}
\label{linearmatrixequation}
    \mb Y = \mb H \mb P + \mb N,
\end{equation} 
where $\mb Y \in \mathbb{C}^{N_r \times B} $ is the received signal matrix, $\mb H \in \mathbb{C}^{N_r \times N_t}$ the channel matrix, $\mb P \in \mathbb{C}^{N_t \times B}$ the pilot matrix, and $\mb N \in \mathbb{C}^{N_r \times B}$ the noise which is uncorrelated with $\mb H$. Vectorizing $\mb Y$ in (\ref{linearmatrixequation}) gives  
\begin{equation}
\label{linearvectorequation}
          \mb y = \widetilde {\mb P}\mb h + \mb n, 
\end{equation}
where $\mb y   = {\rm vec}( \mb Y$), $ \widetilde{\mb P}  =  \mb P^T \otimes {\mb I} $, $\mb h   = {\rm vec}( \mb H$), $\mb n= {\rm vec}( \mb N$), $\rm vec(\cdot)$ denotes vectorization, and $\otimes$ denotes Kronecker product. We assume a Rayleigh fading channel and denote by $\bm \Sigma_{\mb h}  \in \mathbb{C}^{N_tN_r \times N_tN_r}$ the covariance matrix of the channel vector $\mb h$. We also assume that the disturbance $\mb n$ is complex Gaussian-distributed with a zero mean and identity covariance matrix.  

Given $\bm \Sigma_{\mb h}$, the MMSE estimate of $\mb h$ from $\mb y$ can be computed as \cite{Ottersten2010} 
\begin{equation}
\label{MMSE}
\widehat{\mb h}_{\rm MMSE} =\bm \Sigma_{\mb h} \widetilde{\mb P}^\dag ( \widetilde{ \mb P}\bm \Sigma_{\mb h} \widetilde{\mb P}^\dag +\mb I)^{-1}\mb y.  
\end{equation}
The covariance matrix $\bm \Sigma_{\mb h}$, which can be very large, must be estimated in order to compute $\widehat{\mb h}_{\rm MMSE}$. In communication systems, $\mb h$ is not directly observable and thus the SCM estimator can not be directly applied to generate $\bm \Sigma_{\mb h}$.    
One may estimate $\bm \Sigma_{\mb h}$ from least squares (LS) estimates of $\mb H$, i.e.,   
\begin{equation} \label{HLSs}
    \widehat{\mb H}_{\rm LS} = \mb Y \mb P^\dag (\mb P \mb P^\dag)^{-1}. 
\end{equation}    
When orthogonal training signal with $\mb P = \sqrt{P} \mb I$ is applied, where $P$ determines the power for training signals, it can be shown that 
\begin{equation} \label{HLS}
    \widehat{\mb H}_{\rm LS} = \frac{1}{\sqrt{P}} \mb Y  =\frac{1}{\sqrt{P}} (\mb H \mb P + \mb N) = \mb H + \frac{1}{\sqrt{P}} \mb N.   
\end{equation}    
Denote by $\widehat{\mb h}_{\mathrm {LS}}$ the vectorization of $\widehat{\mb H}_{\mathrm {LS}}$. It can be shown that the covariance matrix of $\widehat{\mb h}_{\mathrm {LS}}$  is  
\begin{equation} \label{SigmaHLS}
   \bm \Sigma_{ \widehat{\mb h}_{\rm LS}} \triangleq \mathrm{E}[ \widehat{\mb h}_{\mathrm {LS}}\widehat{\mb h}_{\mathrm {LS}}^\dag ]= \bm \Sigma_{\mb h} + \frac{1}{P} \mb I.    
\end{equation}    
Therefore, if $\bm \Sigma_{ \widehat{\mb h}_{\rm LS}} $ is estimated as $\widehat{\bm \Sigma}_{ \widehat{\mb h}_{\rm LS}}$, we can then use (\ref{SigmaHLS}) to estimate $\bm \Sigma_{\mb h}$ as 
\[
	 \widehat{\bm \Sigma}_{\mb h} = \widehat{\bm \Sigma}_{ \widehat{\mb h}_{\rm LS}} - \frac{1}{P} \mb I,     
\] 
which can be used in (\ref{MMSE}). 
The estimation of $\bm \Sigma_{ \widehat{\mb h}_{\rm LS}} $ can be achieved using the different shrinkage estimators introduced in this paper.

\begin{figure} \centering 
                \label{Result}
                 \includegraphics[width=0.85\columnwidth]	{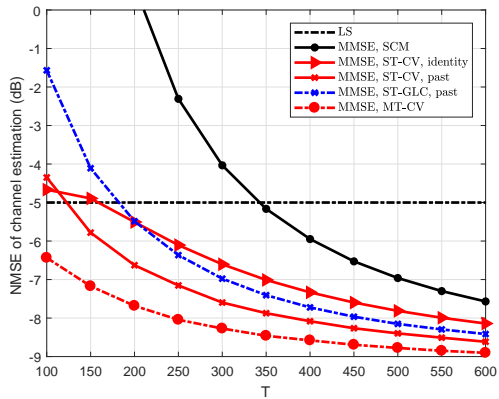}
                 \caption{Performance of MMSE estimation of MIMO channels with the channel covariance matrix estimated using different estimators with $N_t=N_r=B=10$. Pilot-to-noise ratio is $5$ dB. ``LS" refers to the LS estimator in (\ref{HLSs}); ``MMSE" refers to the MMSE channel estimator (\ref{MMSE}) constructed using estimated covariance matrices; ``identity" and ``past" represent shrinkage targets chosen as the scaled identity matrix and the estimate of a past covariance matrix, respectively; ``MT-CV" uses both the identity target and the target set as a past estimate. }  
  \end{figure}  

An example is shown in Fig. 4. The covariance matrix is assumed to be 
\begin{equation}
	\label{Kronecker}
	   \bm \Sigma_{\mb h} =  \bm \Sigma_t   \otimes \bm \Sigma_r,
\end{equation} 
where $\bm \Sigma_t$ and $\bm \Sigma_r$ are, respectively, the transmitter side and receiver side covariance matrix, with entries given by 
\begin{equation} 	\label{Rtt}
	[\bm \Sigma_t]_{i,j} = \left\{ \begin{array}{cc} r_t^{|i-j|}, & i\ge j \\
															{(r_t^\ast)}^{|i-j|}, & i < j 
										  \end{array} \right., 
\end{equation} 
\begin{equation}\label{Rrr}
	[\bm \Sigma_r]_{i,j} = \left\{ \begin{array}{cc} r_r^{|i-j|}, & i\ge j \\
															{(r_r^\ast)}^{|i-j|}, & i < j 
										  \end{array} \right., 
\end{equation} 
$r_t = 0.7 e^{-j 0.9349 \pi}$ and $r_r = 0.9 e^{-j 0.9289 \pi}$. While applying shrinkage to estimate $ \bm \Sigma_{ \widehat{\mb h}_{\rm LS}}$, two shrinkage targets are tested: the identity matrix and the shrinkage estimate (with a scaled identity target) of a past covariance matrix. The second is considered based on the assumption that $\bm \Sigma_{\mb h}$ is slowly varying in time and a well-conditioned estimate of a past covariance matrix $\bm \Sigma^{\mathrm{past}}_{\mb h}$ can be available. In our simulations, $\bm \Sigma^{\mathrm{past}}_{\mb h}$ is modeled by randomly perturbing $r_t$ and $r_r$ in (\ref{Rtt}) and (\ref{Rrr}) by $\delta_{t}$ and $\delta_{r}$ whose real and imaginary parts are both randomly and uniformly generated from $\left[-\frac{1}{10\sqrt{2}}, \frac{1}{10\sqrt{2}}\right ]$. The normalized MSE of channel estimation is defined as 
\begin{equation} 
	\mathrm{NMSE}_{\mb h} \triangleq \frac{\mathrm{E}[ || \widehat{\mb h}_{\rm MMSE}   - \mb h ||_F^2]}{\mathrm{E}[ || \mb h||_F^2]},   
\end{equation} 
where $\widehat{\mb h}_{\rm MMSE}$ is the MMSE channel estimate obtained from (\ref{MMSE}) with the true channel covariance matrix replaced by its shrinkage estimate.      

From the simulation results in Fig. 4, when the number of samples  $T$ of channel estimates is small, the MMSE channel estimator constructed using the SCM estimate of $\bm \Sigma_{\mb h}$ is poorer than the LS estimator which does not require any knowledge of $\bm \Sigma_{\mb h}$. Therefore, an accurate estimate of the covariance matrix is necessary to exploit the potential of the MMSE channel estimator. Shrinkage with LOOCV choice of the shrinkage coefficients  improves the performance of the MMSE channel estimator by providing a better estimate of $\bm \Sigma_{\mb h}$. Two-target shrinkage can further enhance performance.  Note that the multi-target method of \cite{Lancewicki2014} is not directly applicable to the shrinkage target used here.  Similarly to \cite{Shariati2014}, we do not exploit the Kronecker product structure in (\ref{Kronecker}) and the exponential modeling of (\ref{Rtt}) and (\ref{Rrr}) while estimating the covariance matrix and similar trends can be observed when the channel covariance matrix follows different models such as those in \cite{Weichselberger2006, Fang2017}.

\emph{Example 5: Application to LMMSE signal estimation:} 
Another example application is the design of linear minimum mean squared error (LMMSE) estimator \cite{Tse2005, Kim2008} for estimating the transmitted signal $\mb x$ in MIMO communications. The received signal is modeled by (\ref{linearmodel}) and the LMMSE estimate of $\mb x$ is obtained as 
\begin{equation} \label{MIMODetection}
	\widehat{\mb x} =  {\mb H}^\dag \bm \Sigma_{\mb y}^{-1} \mb y, 
\end{equation} 
where we have assumed that $\mb x$ has identity covariance matrix and $\bm \Sigma_{\mb y}$ is the covariance matrix of $\mb y$. The OLS-based covariance matrix estimation in Section \ref{LOOCVOLS} can be used to estimate $\bm \Sigma_{\mb y}$ in (\ref{MIMODetection}).  In Fig. 5, we show an example where the shrinkage target $\mb T_0$ is chosen as the diagonal matrix of the OLS estimate (\ref{ROLS}) of the covariance matrix. This results in a shrunk LMMSE signal estimator
\begin{equation}
	\widehat{\mb x} = \widehat{\mb H}^\dag ( \rho( \widehat{\mb H}  \widehat{\mb H}^\dag  + \widehat{\sigma^2}  \mb I) + \tau \mb T_0 ) ^{-1} \mb y.  
\end{equation} 
Orthogonal training of length $T$ constructed from the discrete Fourier transform (DFT) matrix is assumed for the OLS channel estimate and finding the shrinkage coefficients using our proposed LOOCV method is achieved at a low complexity. The normalized MSE of signal estimation is defined as 
\begin{equation} 
	\mathrm{NMSE}_{\mb x} \triangleq \frac{\mathrm{E}[ || \widehat{\mb x} - \mb x ||_F^2]}{\mathrm{E}[ || \mb x||_F^2]}.    
\end{equation} 
Fig. 5 presents the simulation results  averaged over $1000$ random realizations of $\mb H$ for each $T$. It can be seen that the shrinkage estimate of the covariance matrix can lead to noticeable improvement of the MSE performance of signal estimation. The resulting performance can approach the oracle choice of $(\rho, \tau)$ that minimizes the MSE of estimating $\mb x$ \cite{Tong2016}. Note that in contrast to the cross-validation methods in \cite{Tong2016} and \cite{Tong2016ICASSP} which choose shrinkage factors by a grid search for optimizing the signal estimation performance, the method proposed in this paper has an analytical solution and optimizes covariance matrix estimation. It also differs from \cite{Tong2017} which targets the design of a signal estimator that shrinks the sample LMMSE filter toward the matched filter.

    	\begin{figure} \centering 
                \label{OLSMIMO}
                 \includegraphics[width=0.85\columnwidth]
                 {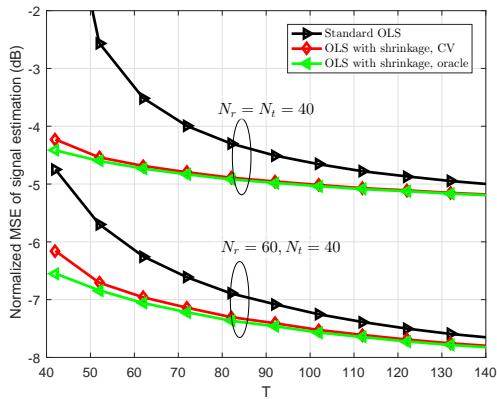}
                 \caption{Performance of the LMMSE signal estimator with channel matrix and received signal's covariance matrices estimated using OLS and shrinkage. The entries of $\mb H$ are independently generated from complex Gaussian distribution with zero mean and variance $1/40$, and the noise variance $\sigma^2 =0.1$.}   
  \end{figure}

\emph{Example 6: Application to MVDR beamforming:}  
Finally, we show an example application to minimum variance distortion-less response (MVDR) beamforming \cite{Mestre06, Montse2014}. We assume a $N=30$-element uniform linear array (ULA) with half-wavelength spacing between neighboring antennas. As in \cite{Montse2014}, we assume that the desired complex Gaussian signal has an angle of arrival (AoA) of $\theta_0=0^{\circ}$ and there are $8$ complex Gaussian interferences in the directions $\{\theta_m\}$ $= \{8^{\circ},$  $-15^{\circ},$ $23^{\circ},$ $-21^{\circ},$ $46^{\circ},$ $-44^{\circ},$ $-85^{\circ},$ $74^{\circ}\}$, all with an average power $10$ dB higher than the desired signal. The noise is assumed to be additive white Gaussian noise (AWGN) with an average power $10$ dB lower than the desired signal. The MVDR beamformer is given by   
\begin{equation} \label{MVDR}
	 {\mb w} =  \frac{   {\mb \Sigma }^{-1}   {\mb s}}{ {\mb s}^\dag   {\mb \Sigma }^{-1}   {\mb s}}, 
\end{equation} 
where $ {\mb s}$ is the steering vector of the desired signal and $ {\mb \Sigma}$ is the covariance matrix of the received signal. We consider a practical scenario where the desired signal's steering vector suffers from an AoA error uniformly distributed in $[-5^\circ, 5^\circ]$ and ${\mb \Sigma}$ is estimated from the training samples by shrinking the SCM $\mb R$ toward the scaled identity matrix $\frac{\mathrm{tr}(\mb R)}{N} \mb I$. We focus on the low-sample-support case and compare the result with an approach that uses the pseudo-inverse of the SCM for computing $\mb w$. The output signal-to-interference-and-noise ratio (SINR) averaged over $1000$ repetitions are plotted in Fig. 6. It is seen that though the proposed approach targets covariance matrix estimation only and is not optimized for beamformer designs, it still provides noticeable gains as compared to the pseudo-inverse approach in the low-sample-support regime.  
    	\begin{figure} \centering 
                \label{MVDRSINR}
                 \includegraphics[width=0.85\columnwidth]
                 {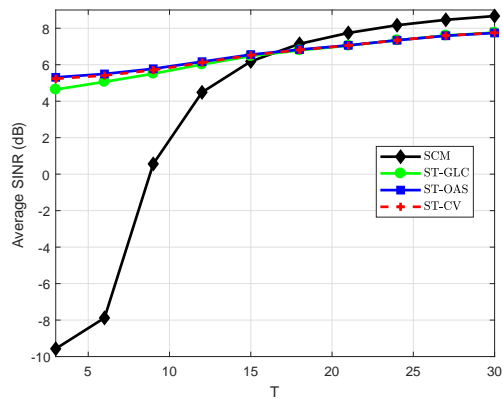}
                 \caption{Average output SINR for a MVDR beamformer with AoA mismatch and the estimated covariance matrix. The results labeled by ``SCM" is obtained by replacing $ {\mb \Sigma }^{-1}$ in (\ref{MVDR}) with the pseudo-inverse of the SCM. Note that the LOOCV and OAS methods achieve almost the same performance, which is slightly better than the GLC method when $T$ is very small.}  
  \end{figure}

\section{Conclusions}
In this paper, we have introduced a leave-one-out cross-validation (LOOCV) method for choosing the coefficients for linear shrinkage covariance matrix estimators. By employing a quadratic loss as the LOOCV prediction error, analytical expressions of the optimal shrinkage coefficients are obtained, which do not require a grid search of the parameters. As a result, the coefficients can be computed at low costs for the SCM- and OLS-based estimation of the covariance matrix. The LOOCV method is generic in the sense that it can be applied to different covariance matrix estimation methods and different shrinkage targets. Numerical examples show that it can approximate the oracle parameter choices in general and have wider applications than several existing analytical methods that have been widely applied.  

Zero-mean signals have been assumed in this paper. When nonzero-mean signals are considered, our proposed approach may be applied after subtracting an estimate of the mean from the samples. However, the inaccuracy in the mean vector estimate may introduce extra errors to the covariance matrix estimation. Jointly estimating the mean and covariance matrix in a robust manner may be further explored. Other future work includes theoretical study of the properties of the proposed approach and low-complexity cross-validation schemes for choosing shrinkage factors for specific signal processing applications such as beamforming, space-time adaptive processing, correlation analysis, etc.

\section*{Acknowledgments}

The authors wish to thank Prof. Antonio Napolitano and the anonymous reviewers for their constructive comments which have greatly improved the paper. This work was supported in part by an International Links grant of University of Wollongong (UOW), Australia, and in part by NSFC under Grant 61601325.  
								
%\section*{References}


\begin{thebibliography}{99}


\bibitem{Scharf91} L. L. Scharf, 
		 \emph{Statistical Signal Processing: Detection, Estimation, and Time Series Analysis}, 
               Addison--Wesley, Bosten, 1991.                 


\bibitem{Ledoit2004}
 		O. Ledoit and M. Wolf, 
		``A well-conditioned estimator for large-dimensional covariance matrices," 
			\emph{J. Multivar. Anal.}, vol. 88, pp. 365-411, 2004.

\bibitem{Stoica2008} 
             P. Stoica, J. Li, X. Zhu, and J. R. Guerci,
		``On using a priori knowledge in space-time adaptive processing,"
		\emph{IEEE Trans. Sig. Process.}, vol. 56, no. 6, pp. 2598-2602, 2008.  

\bibitem{Chen2010} 
		Y. Chen, A. Wiesel,  Y. C. Eldar, and A. O. Hero, 
		``Shrinkage algorithms for MMSE covariance estimation,"
		 \emph{IEEE Trans. Sig. Process.}, vol. 58, no. 10, pp. 5016-5029, 2010. 

\bibitem{Du2010} 
            L. Du, J. Li, and P. Stoica, 
             ``Fully automatic computation of diagonal loading levels for robust adaptive beamforming," 
             \emph{IEEE Trans. Aerosp. Electron. Syst}, vol. 46, no. 1, pp. 449-458, 2010.  

\bibitem{Bickel2008a} 
           P. J. Bickel and E. Levina, ``Regularized estimation of large covariance matrices," \emph{Ann. Statist.}, vol. 36, no. 1, pp. 199-227, 2008. 


\bibitem{Bickel2008b} 
           P. J. Bickel and E. Levina, ``Covariance regularization by thresholding," \emph{Ann. Statist.}, vol. 36, no. 6, pp. 2577-2604, 2008.

\bibitem{Stein} 
           C. Stein, ``Inadmissibility of the usual estimator for the mean of a multivariate normal distribution," \emph{Proc. Third Berkeley Symp. Math. Statist. Prob.}, 1, pp. 197-206, 1956.

\bibitem{Haff1980}
            L. R. Haff, ``Empirical Bayes estimation of the multivariate normal covariance matrix," \emph{Ann. Statist.}, vol. 8, no. 3, pp. 586–597, 1980. 


\bibitem{Hoffbeck1996}
		J. P. Hoffbeck and D.A. Landgrebe, 
            ``Covariance matrix estimation and classification with limited training data," 
             \emph{IEEE Trans. Pattern Anal. Mach. Intell.}, vol. 18, no. 7, pp.763-767, 1996. 

\bibitem{Daniels2001} 
            M. Daniels and R. Kass, ``Shrinkage estimators for covariance matrices," \emph{Biometrics}, vol. 57, pp. 1173-1184, 2001. 
            
\bibitem{Schafer2005} 
              J. Sch\"afer and K. Strimmer, ``A shrinkage approach to large-scale covariance matrix estimation and implications for functional genomics," \emph{Statist. Appl. Genetics Molecular Biol.}, vol. 4, 2005.

\bibitem{Warton2008}
D. I. Warton, ``Penalized normal likelihood and ridge regularization of correlation and covariance matrices," \emph{JASA}, vol. 103, no. 481, pp. 340-49, 2008.  

\bibitem{Fisher2011} T. J. Fisher and X. Sun, 
           ``Improved Stein-type shrinkage estimators for the high-dimensional multivariate normal covariance matrix," \emph{Comp. Statist. Data Analysis}, 55, 1909-1918, 2011. 


\bibitem{Chen12} X. Chen, Z. Jane Wang, and M. J. McKeown, 
             ``Shrinkage-to-tapering estimation of large covariance matrices," \emph{IEEE Trans. Sig. Process.}, vol. 60, pp. 5640-5656, 2012.

\bibitem{Theiler2012} J. Theiler, 
                        ``The incredible shrinking covariance estimator,"  
                            \emph{Proc. SPIE.}, vol. 8391, pp. 83910P, 2012.

\bibitem{Lancewicki2014} 
T. Lancewicki and M. Aladjem, ``Multi-target shrinkage estimation for covariance matrices,"  \emph{IEEE Trans. Sig. Process.}, vol. 62, no. 24, pp. 6380-6390, 2014.

\bibitem{Ikeda2016}
                Y. Ikeda, T. Kubokawa, and M. S. Srivastava, 
                   ``Comparison of linear shrinkage estimators of a large covariance matrix in normal and non-normal distributions," \emph{Comput. Stat. Data Anal.}, vol. 95, 95-108, 2016.


\bibitem{Tong2014} T. Tong, C. Wang, and Y. Wang, 
           ``Estimation of variances and covariances for high-dimensional data: a selective review," 
               \emph{Wiley Interdiscip. Rev. Comput. Stat.}, vol. 6, no. 4, pp. 255-264, 2014.  

\bibitem{Senneret2016} M. Senneret, Y. Malevergne, P. Abry, G. Perrin, and L. Jaffrès, ``Covariance versus precision matrix estimation for efficient asset allocation," 
\emph{IEEE J. Sel. Topics Sig. Process.}, vol. 10, no. 6, pp. 982-993, Sept. 2016. 


\bibitem{Fan2016} J. Fan, Y. Liao, and H. Liu, 
     ``An overview of the estimation of large covariance and precision matrices," 
        \emph{Econom. J.} 19, no. 1, C1-C32, 2016.




\bibitem{Bartz2016} O. Ledoit and M. Wolf, ``Nonlinear shrinkage estimation of large-dimensional covariance matrices," \emph{Ann. Statist.}, vol. 40, no. 2, 1024-1060, 2012. 



\bibitem{Lam2016} 
C. Lam, ``Nonparametric eigenvalue-regularized precision or covariance matrix estimator,"  \emph{Ann. Statist.}, vol. 44, no. 3, pp. 928-953, 2016.
 

 \bibitem{Aubry2012} 
 A. Aubry, A. De Maio, L. Pallotta, and A. Farina, ``Maximum likelihood estimation of a structured covariance matrix with a condition number constraint," \emph{IEEE Trans. Signal Process.}, vol. 60, pp. 3004-3021, 2012.


\bibitem{Won2013}
J.-H. Won, J. Lim, S.-J. Kim, and B. Rajaratnam, ``Condition-number regularized covariance estimation," \emph{J. Roy. Statist. Soc. B}, vol. 75, pp. 427-450, Jun. 2013. 

 
\bibitem{Kourtis2012}

A. Kourtis, G. Dotsis, and R. N. Markellos, ``Parameter uncertainty in portfolio selection: Shrinking the inverse covariance matrix," \emph{J. Bank. Financ.}, vol. 36, no. 9, pp.2522-2531, 2012.



\bibitem{Zhang2013} M. Zhang, F. Rubio, and D. P. Palomar, ``Improved calibration of high-dimensional precision matrices," \emph{IEEE Trans. Sig. Process.}, vol. 61, no. 6, pp. 1509-1519,  2013.
 


\bibitem{Wang2015} 
C. Wang, G. Pan, T. Tong, and L. Zhu, ``Shrinkage estimation of large dimensional precision matrix using random matrix theory," \emph{Statistica Sinica}, vol. 25, no. 3, pp. 993-1008, 2015.


\bibitem{Ito2015}
T. Ito and T. Kubokawa, ``Linear ridge estimator of high-dimensional precision matrix using random matrix theory," \emph{Technical Report F-995, CIRJE}, Faculty of Economics, University of Tokyo, 2015.  

\bibitem{Bodnar2016} T. Bodnar, A. K. Gupta, and N. Parolya, 
``Direct shrinkage estimation of large dimensional precision matrix," 
\emph{J. Multivar. Anal.}, vol. 146, pp. 223-236, 2016. 

\bibitem{Mestre06} X. Mestre and M. A. Lagunas, 
		``Finite sample size effect on minimum variance beamformers: Optimum diagonal loading factor for large arrays,"  
            	  	\emph{IEEE Trans. Sig. Process.},  
            	  	vol. 54, no. 1, pp. 69-82, 2006.  

\bibitem{Wen13} 
               C.-K. Wen, J.-C. Chen, and P. Ting, 
               ``A shrinkage linear minimum mean square error estimator," 
               \emph{IEEE Sig. Process. Lett.,} vol. 20, no. 12, pp.1179-1182, 2013.           
  
 

\bibitem{Montse2014} 
		  J. Serra and M. N\'ajar, 
		  	``Asymptotically optimal linear shrinkage of sample LMMSE and MVDR filters," 
		  	\emph{IEEE Trans. Sig. Process.},  
		  	vol. 62, no. 14, pp. 3552-3564, 2014.  


\bibitem{Zhang13}
		M. Zhang, F. Rubio, D. Palomar, and X. Mestre,
			``Finite-sample linear filter optimization in wireless communications and financial systems,''
		\emph{IEEE Trans. Sig. Process.,}
			vol. 61, no. 20, pp. 5014-5025, 2013.




\bibitem{Tong2016}  {J. Tong}, P. J. Schreier, Q. Guo,  S. Tong, J. Xi, and Y. Yu,   
			``Shrinkage of covariance matrices for linear signal estimation using cross-validation," 
                            \emph{IEEE Trans. Sig. Process.}, vol. 64, no. 11, pp. 2965-2975, 2016.   
 
 
\bibitem{Tong2016ICASSP}   J. Tong, Q. Guo, J. Xi, Y. Yu, and P. J. Schreier, 
``Choosing the diagonal loading factor for linear signal estimation using cross validation," in \emph{Proc. IEEE ICASSP 2016},  pp. 3956-3959, 2016. 


\bibitem{Guerci2006} 

J. R. Guerci and E. J. Baranoski, 
``Knowledge-aided adaptive radar at DARPA: an overview," 
\emph{IEEE Signal Process. Mag.}, vol. 23, no. 1, pp. 41-50, Jan. 2006. 


 \bibitem{Arlot10}  S. Arlot and A. Celisse, 
		``A survey of cross-validation procedures for model selection," 
		 \emph{Statist. Surv.}, vol. 4, pp. 40-79, 2010. 


\bibitem{GCV79} 
G. H. Golub,  M. Heath, and G. Wahba, 
``Generalized cross-validation as a method for choosing a good ridge parameter,"    \emph{Technometrics}, vol. 21, no. 2, pp. 215-223, 1979. 


\bibitem{Nowak97}
		 R. D. Nowak, 
		 ``Optimal signal estimation using cross-validation,"
		 \emph{IEEE Sig. Process. Letters,} vol. 4, no. 1, pp. 23-25, 1997. 


\bibitem{Ottersten2010} 
              E. Bj\"ornson and  B.  Ottersten, ``A  framework  for  training-based  estimation  in  arbitrarily  correlated  Rician  MIMO  channels  with  Rician disturbance," \emph{IEEE  Transactions  on  Signal  Processing,} vol.  58,  no.  3, pp. 1807–1820, March 2010. 

\bibitem{Shariati2014} 
N. Shariati, E. Bj\"ornson, M. Bengtsson and M. Debbah, 
``Low-complexity polynomial channel estimation in large-scale MIMO with arbitrary statistics," \emph{IEEE J. Sel. Topics Signal Process.}, vol. 8, no. 5, pp. 815-830, Oct. 2014. 


\bibitem{Weichselberger2006} W. Weichselberger, M. Herdin, H. Ozcelik, and E. Bonek, 
``A stochastic MIMO channel model with joint correlation of both link ends," \emph{IEEE Trans. Wireless Commun.,} vol. 5, no. 1, pp. 90-100, 2006. 

\bibitem{Fang2017} 
J. Fang, X. Li, H. Li and F. Gao, ``Low-rank covariance-assisted downlink training and channel estimation for FDD massive MIMO systems," 
\emph{IEEE Trans. Wireless Commun.}, vol. 16, no. 3, pp. 1935-1947, March 2017.

\bibitem{Tse2005} D. Tse and P. Viswanath, \emph{Fundamentals of Wireless Communications.} Cambridge, U.K.: Cambridge Univ. Press, 2005.

\bibitem{Kim2008}
N. Kim, Y. Lee and H. Park, 
``Performance analysis of MIMO system with linear MMSE receiver," 
\emph{IEEE Trans. Wireless Commun.}, vol. 7, no. 11, pp. 4474-4478, Nov. 2008. 


\bibitem{Tong2017} 
J. Tong, J. Xi, Q. Guo and Y. Yu, ``Low-complexity cross-validation design of a linear estimator," \emph{Electronics Letters,} vol. 53, no. 18, pp. 1252-1254, 2017. 


\end{thebibliography}
\end{document}